\documentclass[11pt]{article}
\PassOptionsToPackage{breaklinks}{hyperref}

\usepackage[hyperref]{emnlp-ijcnlp-2019}
\usepackage{times}
\usepackage{latexsym}

\usepackage{graphicx}
\usepackage{multirow}
\usepackage{multicol}
\usepackage{amssymb}
\usepackage{wasysym}
\usepackage{arydshln}

\aclfinalcopy

\newcommand{\Sec}[1]{Section~\ref{sec:#1}}
\newcommand{\Fig}[1]{Figure~\ref{fig:#1}}
\newcommand{\Tab}[1]{Table~\ref{tab:#1}}

\title{Hybrid Data-Model Parallel Training for Sequence-to-Sequence\\ Recurrent Neural Network Machine Translation\thanks{\, The contents in this manuscript are identical to those in our formal publication at the 8th Workshop on Patent and Scientific Literature Translation (PSLT 2019), whereas the style is slightly modified.}}

\author{Junya Ono \qquad Masao Utiyama \qquad Eiichiro Sumita \\
	National Institute of Information and Communications Technology \\
	3-5 Hikaridai, Seika-cho, Soraku-gun, Kyoto, 619-0289, Japan\\
	\{\tt junya.ono, mutiyama, eiichiro.sumita\}@nict.go.jp
}
\date{}

\begin{document}
\maketitle

\begin{abstract}
\normalsize
Reduction of training time is an im-portant issue in many tasks like patent translation involving neural networks. Data parallelism and model parallelism are two common approaches for reduc-ing training time using multiple graphics processing units (GPUs) on one machine. In this paper, we propose a hybrid data-model parallel approach for sequence-to-sequence (Seq2Seq) recurrent neural network (RNN) ma-chine translation. We apply a model parallel approach to the RNN encoder-decoder part of the Seq2Seq model and a data parallel approach to the attention-softmax part of the model. We achieved a speed-up of 4.13 to 4.20 times when using 4 GPUs compared with the training speed when using 1 GPU without affecting machine translation accuracy as measured in terms of BLEU scores.
\end{abstract}

\section{Introduction}
\label{sec:intro}

Neural machine translation (NMT) has been widely used owing to its high accuracy. A downside of NMT is it requires a long training time. For instance, training a Seq2Seq RNN machine translation (MT) with attention \cite{Luong:15} could take over 10 days using 10 million sentence pairs.
A natural solution to this is to use multiple GPUs. There are currently two common approaches for reducing the training time of NMT models. One approach is by using data parallel approach, while the other approach is through the use of the model parallel approach.

The data parallel approach is common in many neural network (NN) frameworks. For instance, OpenNMT-lua \cite{Klein:17},\footnote{\url{https://github.com/OpenNMT/OpenNMT}} an NMT toolkit, uses multiple GPUs in training NN models using the data parallel approach. In this approach, the same model is distributed to different GPUs as replicas, and each replica is updated using different data. Afterward, the gradients obtained from each replica are accumulated, and parameters are updated.

The model parallel approach has been used for training a Seq2Seq RNN MT with attention \cite{Wu:16}. In this approach, the model is distributed across multiple GPUs, that is, each GPU has only a part of the model. Subsequently, the same data are processed by all GPUs so that each GPU estimates the parameters it is responsible for.

In this paper, we propose a hybrid data-model parallel approach for Seq2Seq RNN MT with attention. We apply a model parallel approach to the RNN encoder-decoder part of the Seq2Seq model and a data parallel approach to the attention-softmax part of the model.

The structure of this paper is as follows. In \Sec{relwork}, we describe related work. In \Sec{model}, first, we discuss the baseline model with/without data/model parallelism. Afterward, we present the proposed hybrid data-model parallel approach. In \Sec{expr}, we present a comparison of these parallel approaches and demonstrate the scalability of the proposed hybrid parallel approach. \Sec{conclusion} presents the conclusion of the work.

\section{Related Work}
\label{sec:relwork}

The accuracy of NN models improves as the model sizes and data increases. Thus, it is necessary to use multiple GPUs when training NN models within a short turnaround time.

There are two common approaches for using multiple GPUs in training. One is data parallelism, involving sending different data to different GPUs with the replicas of the same model. The other is model parallelism, involving sending the same data to different GPUs having different parts of the model.

\subsection{Data parallelism}
\label{sec:data_parallelism}

In this approach, each GPU has a replica of the same NN model. The gradients obtained from each model on each GPU are accumulated after a backward process, and the parameters are synchronized and updated.

The advantage of using this model is that it can be applied to any NN model because it does not depend on the model structure. In particular, it can be applied to many models such as Seq2Seq RNN and Inception Network \cite{Abadi:16}. Many deep neural network (DNN) frameworks implement data parallelism.

While data parallelism is general and powerful, it is subject to synchronization issues among multiple GPUs as the model size or the number of model parameters increases. Note that when using multiple machines, asynchronous updates may be used in reducing synchronization costs. However, we focus on using multiple GPUs on one machine, where synchronous updates are generally better than asynchronous updates.

To reduce the synchronization costs relative to all training costs, it is necessary to train models using a large mini-batch size. However, the mini-batch size is bounded by the GPU memory. Furthermore, large mini-batch sizes in general, make convergence difficult and can worsen accuracy of the tasks \cite{Krizhevsky:14,Keskar:17}.

Another important factor to be considered is the ratio of processing time needed for synchronization and forward-backward process on each GPU. If synchronization takes much longer than the forward-backward process, the advantage of using multiple GPUs diminishes.

In summary, depending on models, data parallelism may not work effectively. In such a case, there are methods that can be used to achieve synchronization after several mini-batches or to overlap backward and synchronization process at the same time \cite{Ott:18}. However, these advanced synchronization methods are out of the scope of this study.

\subsection{Model parallelism}
\label{sec:model_parallelism}

In this approach, each GPU has different parameters (and computation) of different parts of a model. Most of the communication occurs when passing intermediate results between GPUs. In other words, multiple GPUs do not need to synchronize the values of the parameters.

In contrast to data parallelism, most DNN frameworks do not implement automatic model parallelism. Programmers have to implement it depending on the model and available GPUs.

Model parallelism needs special care when assigning different layers to different GPUs. For example, each long short-term memory (LSTM) layer may be placed on each GPU in case of stacked-LSTMs in encoder-decoder NN. \citet{Wu:16} have already proposed similar model parallelism for Seq2Seq RNN MT, although they did not describe the actual speed-up achieved.

The scalability of model parallelism is better than that of data parallelism when it works effectively. In data parallelism, when we increase the number of samples in each mini-batch to N times, we expect less than N times speed-up due to synchronization costs.

In contrast, we can expect more than N times speed-up when using model parallelism, owing to the following two reasons. First, we can increase the mini-batch size as in the case of data parallelism. Second, each GPU is able to compute different layers of the model without requiring synchronization.

\subsection{Automatic hybrid parallelism, distributed training, and Transformer}
\label{sec:automatic}

While we focus on hybrid data-model parallelism for Seq2Seq RNN MT in this paper, \citet{Wang:18} have proposed an approach for automatically conducting hybrid data-model parallelism. Applying their method to Seq2Seq RNN MT would be the focus of our future work.

While we focus on parallelism on one machine in this paper, using multiple machines is also a good way of achieving a short turnaround time in training. \citet{Ott:18} reported that a significant speed-up can be obtained while maintaining translation accuracy using data parallelism on 16 machines.

While the Transformer model has recently been demonstrated to have a superior translation performance to the Seq2Seq RNN MT with attention \cite{Vaswani:17}, we focus on how to combine data parallelism and model parallelism in Seq2Seq RNN MT with attention. We believe the proposed hybrid parallel approach to be applicable to the Transformer translation model because Transformer also has an encoder, decoder, and softmax layers. However, we would leave the application of the proposed hybrid data-model parallel approach to Transformer as a part of our future work.

\section{Model Structure and Parallelism}
\label{sec:model}

\subsection{Baseline model}
\label{sec:baseline}

Attention-based NMT has improved translation accuracy compared with the sequence-to-sequence NMT without attention model \cite{Bahdanau:15,Luong:15}.

\begin{figure}[t]
  \centering
  \includegraphics*[width=\columnwidth]{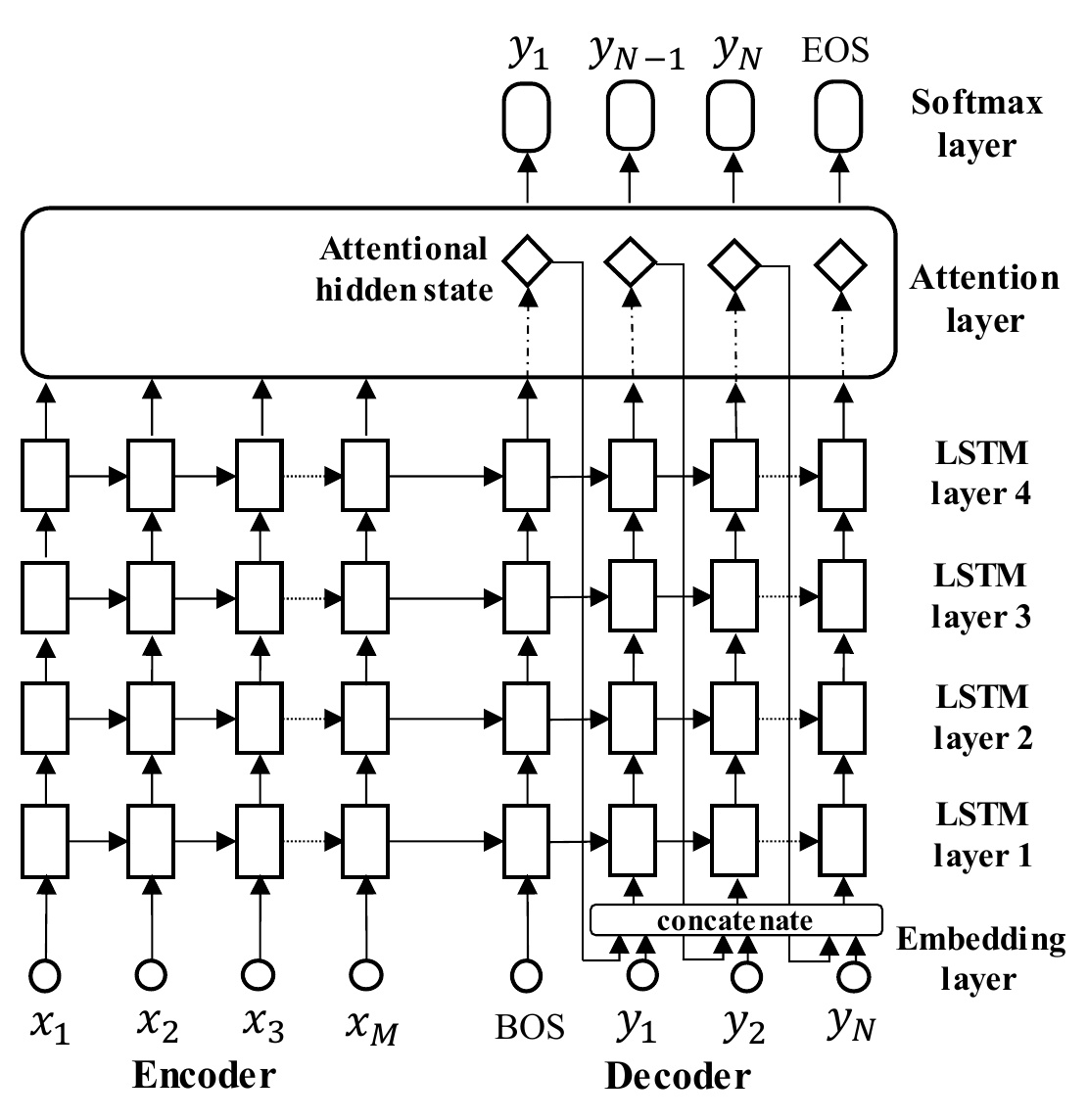}
  \caption{Our baseline model, the attention-based encoder-decoder model \cite{Luong:15}. This model consists of stacked-LSTMs containing 4 layers with the input-feeding approach. The hidden state of attention is concatenated with the target word embedding be-fore being input into the first LSTM layer.}
  \label{fig:figure1}
\end{figure}

\Fig{figure1} shows our baseline model \cite{Luong:15}. The decoder side of this model uses the input-feeding approach, where the hidden state of attention is concatenated with the target word embedding before being input into the first LSTM layer.

Data parallelism can be applied to this baseline model easily. We place each replica of this model on each GPU. Next, the input parallel texts are distributed equally to different GPUs. Finally, synchronization of parameter values is conducted after each forward-backward process.

\begin{figure}[t]
  \centering
  \includegraphics*[width=\columnwidth]{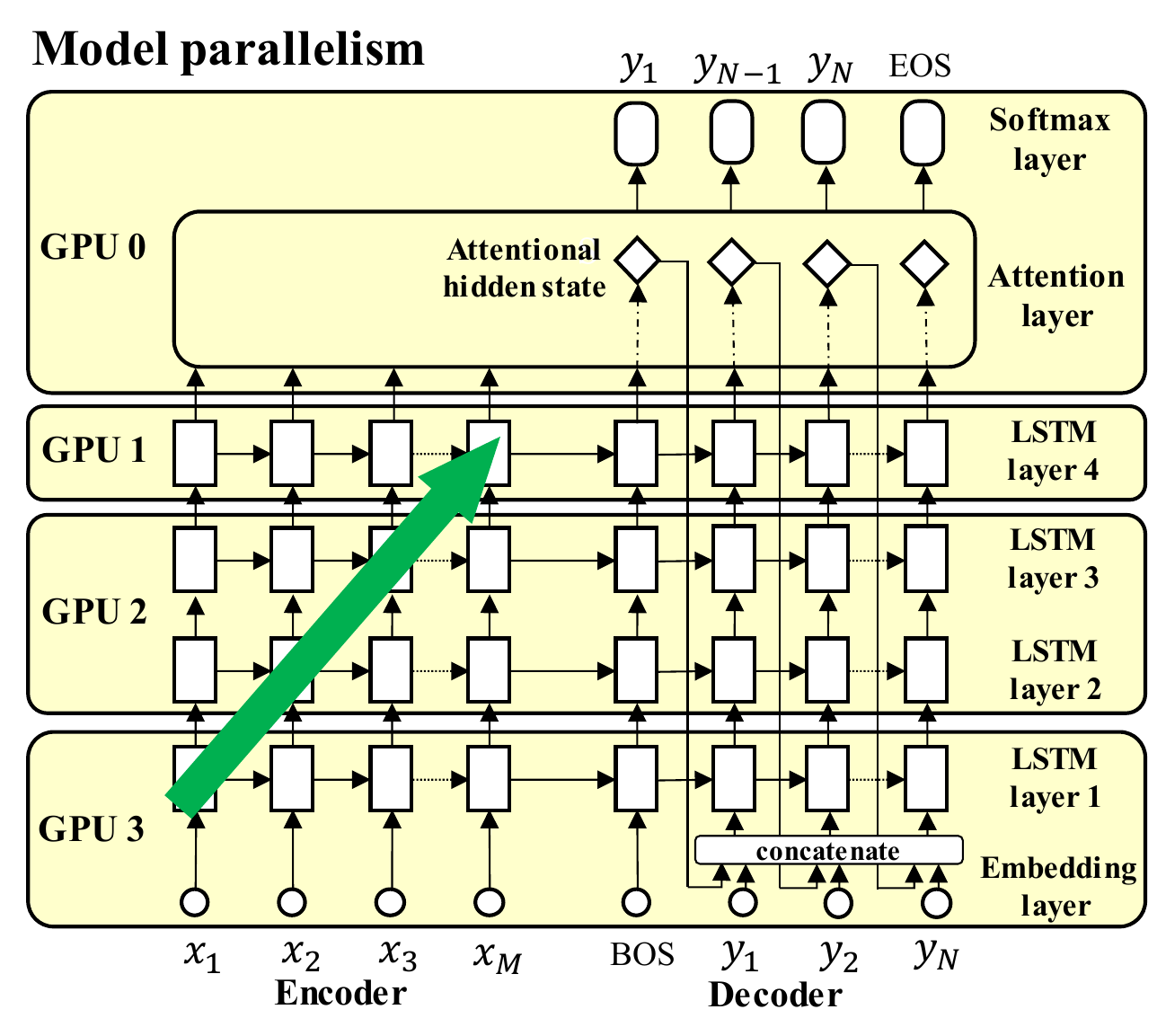}
  \caption{Model parallelism on 4 GPUs for the baseline model of \Fig{figure1}. The same depth layer in the encoder-decoder part is placed on the same GPU. The encoder side allows efficient parallelism, while the decoder part does not due to input-feeding.}
  \label{fig:figure2}
\end{figure}

\Fig{figure2} shows an application of model parallelism to the baseline model on 4 GPUs. In the figure, we assign different layers in the encoder-decoder part to different 3 GPUs. We also assign the attention and softmax layers to 1 GPU. This assignment is based on the fact that the attention-softmax part requires a relatively large GPU memory.

The model parallel approach is effective in this case because there are many parameters in the attention-based encoder-decoder model. Let $U$ be a certain value representing the number of parameters, the embedding layer has $2U$ parameters, each LSTM layer has $8U$ parameters (a total of $32U$ parameters), and the attention-softmax part has $4U$ parameters. When using model parallelism, it is not necessary to synchronize these parameters. We only have to pass intermediate results between different GPUs.

Note that the green arrow in \Fig{figure2} is pointing to the upper right direction. It indicates that the computation of one node can start immediately after the left and down nodes finish their computation. In this way, in the encoder side, GPUs can work without waiting for the completion of the computation in the previous steps.

In contrast, the nodes in the decoder side cannot start performing their assigned computations until all nodes related to the previous target words finish their computation. This is due to the input-feeding approach employed. For instance, the target word embedding of $y_2$ needs to be concatenated with the attentional hidden state of $y_1$ before being input into the first LSTM layer.

\subsection{Proposed model for hybrid parallelism}
\label{sec:proposed}

Herein, we propose our hybrid parallelism for Seq2Seq RNN MT. First, we remove input-feeding in the decoder side of the baseline model, and then we introduce our hybrid parallelism.

\begin{figure}[t]
  \centering
  \includegraphics*[width=\columnwidth]{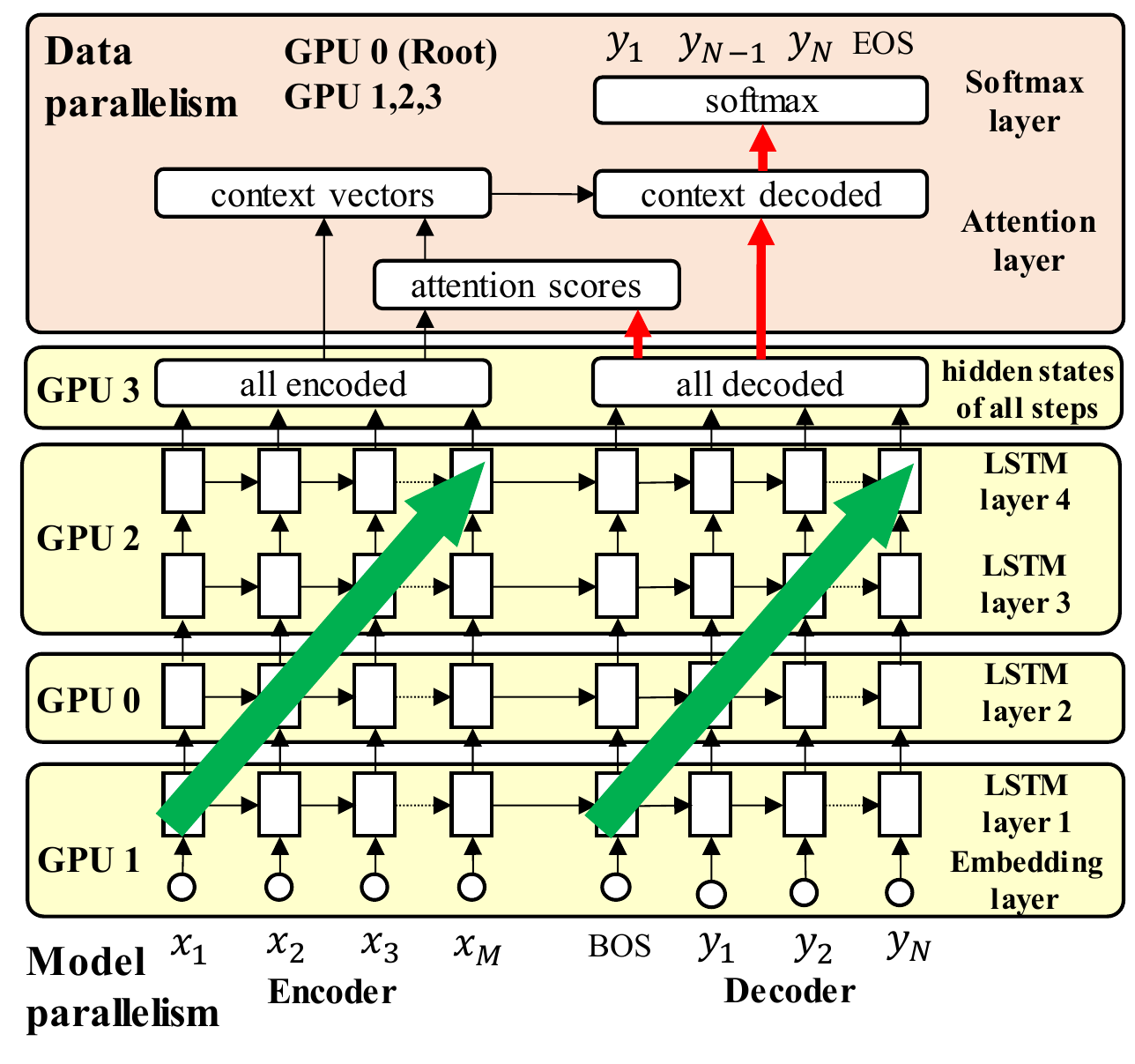}
  \caption{Proposed model for hybrid parallelism.}
  \label{fig:figure3}
\end{figure}

\Fig{figure3} shows our model for hybrid parallelism. First, we employ model parallelism in calculating the states of hidden nodes for all steps in both encoder and decoder sides. Afterward, we apply data parallelism in calculating attention scores, context vectors, and softmax for getting target words.  Note that this is possible because all target words are given beforehand in the training phase.

As stated earlier, we remove input-feeding in the decoder of the baseline model \cite{Luong:15}. While input-feeding has been proposed by \citet{Luong:15} and has shown its advantages in translation accuracy, it has been found to be unsuitable for parallelism. Removing input-feeding removes the dependency of calculation on previous steps in the decoder side. The green arrows going to the upper-right direction show that the computation of a node can start immediately after completion of left and down nodes computation in both encoder and decoder sides. By comparing Figures~\ref{fig:figure2} with \ref{fig:figure3}, we observe that removing input-feeding allows model parallelism to perform better parallel computation. Note that the proposed NMT model has already been proposed by \citet{Luong:15} as a simpler model than the baseline model with input-feeding. However, in the section on the experimentation, we show that removing input-feeding does not affect translation accuracy in terms of BLEU scores obtained.

We now present how we alternate model parallelism and data parallelism on the same 4 GPUs. This is the most important point in the proposed hybrid parallelism implementation.

First, we use 4 GPUs for model parallelism. The source and target word embedding layers and 4 LSTM layers are placed on 3 GPUs as shown in \Fig{figure3}. The remaining GPU (GPU 3 in \Fig{figure3}) stores the hidden states of all steps in the encoder-decoder part.

After the forward process of all hidden states, we move to data parallelism. The intermediate results of all hidden states for all data in the mini-batch are distributed equally to 4 GPUs. While all GPUs have replicas of the same network structure, as shown in \Fig{figure3}, we use GPU 0 as the root for accumulating and synchronizing all parameter values relating to the calculation of attention scores, context vectors, softmax, and so on. The alternation of data parallelism and model parallelism on the backward process goes in a similar but opposite direction.

As mentioned in \Sec{baseline}, the encoder-decoder part has much more parameters than the attention-softmax part. This is the reason why we use model parallelism on the encoder-decoder part and data parallelism on the attention-softmax part.

We now describe closely how we obtain the attention scores and so on in \Fig{figure3}. We omit an explanation of model parallelism for stacked-LSTM layers because it is straightforward.

Let $\alpha$ be ``attention score'' in \Fig{figure3}, it is a concatenation of all attention coefficients of all decoder steps. We employ the attention coefficient defined as global attention \cite{Luong:15}.
\begin{equation}
  \mathbf{\alpha} = (a_{1},\ldots,\alpha_{i},\ldots,\alpha_{N}) = \mathit{Softmax}(\hat{\mathbf{\alpha}})\\
\end{equation}
\begin{equation}
  \hat{\alpha} = \mathbf{H}^{T}W_{\alpha}\mathbf{S}
\end{equation}
where $\mathbf{S}=(S_{1},\ldots,S_{j},\ldots,S_{M})$ denotes the concatenation of all hidden states of length $M$ in the encoder side, $\mathbf{H}=(H_{1},\ldots,H_{i},\ldots,H_{N})$ denotes the concatenation of all hidden states of length $N$ in the decoder side, and $W_{\alpha}$ denotes a parameter matrix. Note that we can calculate $\alpha$ at once after obtaining the hidden states of all steps in the encoder-decoder part in the forward process.

The ``context vectors'' $C$ in \Fig{figure3} can be defined as
\begin{equation}
  \mathbf{C}=(C_{1},\ldots,C_{i},\ldots,C_{N})=\alpha\cdot\mathbf{S}
\end{equation}
The ``context decoded'' $\mathbf{H}_{c}$ in \Fig{figure3} can be defined as
\begin{eqnarray}
  \mathbf{H}_{c} &= &(H_{c1},\ldots,H_{ci},\ldots,H_{cN})\\
  &= &\mathit{tanh}(W_{c}[\mathbf{H};\mathbf{C}])\nonumber
\end{eqnarray}
where $W_{c}$ denotes a parameter matrix. Finally, the conditional probabilities $\mathbf{P}$ of the target sentence words can be computed as
\begin{equation}
  P=(P_{1},\ldots,P_{i},\ldots,P_{N})=\mathit{Softmax}\{F_{c}(\mathbf{H}_{c})\}
\end{equation}
\begin{equation}
  P_{i}=P(y_{i}|y_{1},\ldots,y_{i-1},\mathbf{x})
\end{equation}
where $F_{c}$ denotes a liner function; $\mathbf{x}$ denotes the source sentence in the encoder side; $\mathbf{y}=(y_{1},\ldots,y_{N})$ represents the target sentence in the decoder side.

\section{Experiments}
\label{sec:expr}

We evaluate training speed, convergence speed, and translation accuracy to compare the performance of the proposed approach as shown in \Fig{figure3} (hereafter referred to as HybridNMT) with the baseline model shown in \Fig{figure1} with/without data/model parallelism. We also augment the proposed approach in \Fig{figure3} with input-feeding (hereafter referred to as HybridNMTIF). HybridNMTIF lacks the parallelism in the decoder side but has input-feeding. Consequently, comparing HybridNMT with HybridNMTIF clarifies the advantages of the proposed hybrid parallelism.

\subsection{Data statistics}
\label{sec:data}

We used datasets of WMT14 \cite{Bojar:14}\footnote{\url{http://www.statmt.org/wmt14/translation-task.html}} and WMT17 \cite{Bojar:17}\footnote{\url{http://www.statmt.org/wmt17/translation-task.html}} English--German shared news translation tasks in the experiments. Both datasets were pre-processed using the scripts of the Marian toolkit \cite{Junczys-Dowmunt:18}.\footnote{\url{https://github.com/marian-nmt/marian-examples/tree/master/wmt2017-uedin}} \Tab{table1} shows the number of sentences in these datasets. For the WMT17 dataset, first, we duplicated the provided parallel corpus, and then we augmented the parallel corpus with the pseudo-parallel corpus obtained using back-translation \cite{Sennrich:16:a} of the provided German monolingual data of 10 million (M) sentences. Overall, we used 19 M sentence pairs in the training. We also used the word vocabulary of 32 thousand (K) types from joint source and target byte pair encoding (BPE) \cite{Sennrich:16:b}.

\begin{table}[t]
  \small
  \centering
  \begin{tabular}{l|rr}
    \hline
    \textbf{Dataset} &\multicolumn{2}{c}{\textbf{Sentences}}\\\cline{2-3}
    \textbf{en-de} &\textbf{WMT14} &\textbf{WMT17}\\
    \hline
    Training (original) &4492K &4561K\\
    Training (monolingual) &-- &10000K\\
    Training (all) &4492K &19122K\\
    Development &3000 &2999\\
    Test &3003 &3004\\
    \hline
  \end{tabular}
  \caption{Datasets of WMT14 and WMT17.}
  \label{tab:table1}
\end{table}

\begin{table}[t]
  \small
  \centering
  \begin{tabular}{l|l}
    \hline
    \textbf{Parameter} &\textbf{Value}\\
    \hline
    word embedding size &512\\
    RNN cell type &Stacked-LSTMs\\
    hidden state size &1024\\
    encoder/decoder depth &4\\
    attention type &global\\
    optimizer &Adam\\
    initial learning rate &0.001\\
    learning rate decay &0.7\\
    \hline
  \end{tabular}
  \caption{Model parameters.}
  \label{tab:table2}
\end{table}

\subsection{Parameter settings}
\label{sec:parameter}

Both the baseline model and HybridNMT are trained with the same hyperparameters, as shown in \Tab{table2}. To prevent over-fitting, we set a dropout of 0.3 \cite{Srivastava:14} and used Adam \cite{Kingma:15} of the following setting: $\beta_{1}=0.9$, $\beta_{2}=0.999$, and $\epsilon=1e-8$.

All models were subject to the same decay schedule of learning rate because the convergence speed generally depends on it. In this experiment, the learning rate was multiplied by a fixed value of 0.7 when the perplexity of the development data increased in a fixed interval; an interval of 5,000 and 20,000 batches for WMT14 and WMT17, respectively, reflecting the difference in the number of sentences in these training data.

\begin{table*}[t]
  \small
  \centering
  \begin{tabular}{l|cc|cc|cc}
    \hline
    &\multicolumn{2}{c|}{SRC token / sec}
    &\multicolumn{2}{c|}{Scaling factor}
    &\multicolumn{2}{c}{Mini-batch size}\\
    &WMT14 &WMT17 &WMT14 &WMT17 &WMT14 &WMT17\\
    \hline

    OpenNMT-lua & & & & & &\\
    \quad baseline (1 GPU) &2979 &2757 &1.00 &1.00 &64 &64\\
    \quad\quad w/ data parallelism &4881 &4715 &1.64 &1.71 &256 &256\\
    \hline\hline

    Our implementation & & & & & &\\
    \quad baseline (1 GPU) &2826 &2550 &1.00 &1.00 &64 &64\\
    \quad\quad w/ data parallelism &4515 &4330 &1.60 &1.70 &256 &256\\
    \quad\quad w/ model parallelism &6570 &6397 &2.32 &2.51 &224 &224\\
    \quad HybridNMTIF &9688 &9109 &3.43 &3.57 &224 &224\\
    \quad \textbf{HybridNMT} &\textbf{11672} &\textbf{10716} &\textbf{4.13} &\textbf{4.20} &224 &224\\
    \hline
  \end{tabular}
  \caption{Results of training speed and scaling factors.}
  \label{tab:table3}
\end{table*}

\begin{table*}[t]
  \small
  \centering
  \begin{tabular}{c|c@{~~~}c@{~~~}c@{~~~}c@{~~~}c@{~~~}c|c@{~~~}c@{~~~}c@{~~~}c@{~~~}c@{~~~}c}
    \hline

    OpenNMT-lua
    &\multicolumn{6}{c|}{WMT14 development (test2013)}
    &\multicolumn{6}{c}{WMT17 development (test2016)}\\
    (length,coverage)$\backslash b$
    			&3	&6	&9	&12	&15	&18	&3	&6	&9	&12	&15	&18\\\hline
	(1.0, 0.0)	&21.80 	&21.83 	&21.81 	&21.74 	&21.65 	&21.54 	&31.70 	&31.86 	&31.73 	&31.73 	&31.65 	&31.55\\
	(0.8, 0.0)	&21.80 	&21.80 	&21.77 	&21.71 	&21.60 	&21.47 	&31.70 	&31.85 	&31.73 	&31.71 	&31.62 	&31.53\\
	(0.6, 0.0)	&21.77 	&21.77 	&21.69 	&21.63 	&21.50 	&21.37 	&31.68 	&31.81 	&31.72 	&31.68 	&31.57 	&31.48\\
	(0.4, 0.0)	&21.77 	&21.75 	&21.66 	&21.58 	&21.44 	&21.31 	&31.68 	&31.79 	&31.67 	&31.61 	&31.49 	&31.40\\
	(0.2, 0.0)	&21.77 	&21.75 	&21.65 	&21.56 	&21.42 	&21.28 	&31.65 	&31.79 	&31.64 	&31.59 	&31.48 	&31.38\\
	(0.0, 0.0)	&21.75 	&21.73 	&21.65 	&21.54 	&21.40 	&21.27 	&31.63 	&31.75 	&31.60 	&31.57 	&31.44 	&31.36\\
	(0.2, 0.2)	&21.14 	&21.08 	&21.18 	&21.12 	&21.10 	&21.15 	&30.87 	&30.94 	&30.84 	&30.85 	&30.79 	&30.70\\

    \hline\hline

    HybridNMT
    &\multicolumn{6}{c|}{WMT14 development (test2013)}
    &\multicolumn{6}{c}{WMT17 development (test2016)}\\
    length$\backslash b$
    			&3	&6	&9	&12	&15	&18	&3	&6	&9	&12	&15	&18\\\hline
	1.0 	&22.43 	&22.75 	&22.72 	&22.75 	&22.79 	&22.75 	&32.23 	&32.60 	&32.61 	&32.73 	&32.65 	&32.60\\
	0.8 	&22.43 	&22.71 	&22.63 	&22.67 	&22.67 	&22.63 	&32.20 	&32.52 	&32.59 	&32.70 	&32.67 	&32.62\\
	0.6 	&22.35 	&22.62 	&22.56 	&22.55 	&22.54 	&22.50 	&32.16 	&32.44 	&32.51 	&32.56 	&32.55 	&32.49\\
	0.4 	&22.29 	&22.50 	&22.43 	&22.38 	&22.40 	&22.35 	&32.10 	&32.32 	&32.36 	&32.38 	&32.38 	&32.32\\
	0.2 	&22.26 	&22.37 	&22.29 	&22.24 	&22.26 	&22.20 	&32.02 	&32.19 	&32.25 	&32.26 	&32.21 	&32.16\\
	0.0 	&22.23 	&22.27 	&22.14 	&22.11 	&22.13 	&22.04 	&32.01 	&32.11 	&32.18 	&32.16 	&32.09 	&31.98\\

    \hline
  \end{tabular}
  \caption{BLEU scores obtained using different hyperparameters for WMT14 and WMT17 development data. The upper half shows the results obtained by OpenNMT-lua whereas the lower half is for the proposed HybridNMT. Rows show the different parameters used for normalization. ``$b$'' stands for the beam size.}
  \label{tab:table4}
\end{table*}

The machine type used for training had 4 GPUs of NVIDIA Tesla V100 and was capable of performing direct data transfer among all GPUs using NVLink. We implemented the baseline model with/without data/model parallelism, HybridNMT, and HybridNMTIF in MXNet v1.3.0 \cite{Chen:15}.\footnote{\url{https://github.com/apache/incubator-mxnet}} We also used OpenNMT-lua v0.9.2 \cite{Klein:17} for comparing the models because it implements the baseline model with/without data parallelism. We used the default synchronous mode in OpenNMT-lua and the SGD optimizer as the default settings of the OpenNMT-lua.

\subsection{Comparison of training speed}
\label{sec:comparison1}

\Tab{table3} summarizes the main results of our experiment. In \Tab{table3}, ``SRC tokens / sec'' indicates the number of source tokens processed in one second. This is a standard measure for evaluating training speed; it is also implemented in OpenNMT-lua. ``Scaling factor'' stands for the ratio of ``SRC tokens / sec'' against that of one GPU. The mini-batch sizes were determined by the available GPU memories. Note that mini-batch sizes were about 4 times when using 4 GPUs compared with those obtained when using 1 GPU.

First, the scaling factors of HybridNMT were higher than those of data/model parallelism. They were 4.13 and 4.20 for WMT14 and WMT17 datasets, respectively. This indicates that our hybrid parallel method for Seq2Seq RNN MT is faster than only data/model parallel approaches. Note also that these scaling factors were higher than the number of GPUs (4). This demonstrates the effectiveness of the proposed hybrid parallelism.

Second, the processing speed and scaling factors of OpenNMT-lua and those obtained from our implementation were similar. \Tab{table4} shows that BLEU scores are comparable. These indicate that our implementation is appropriate.

Third, the scaling factors of model parallelism were better than those of data parallelism were. For WMT14, the scaling factor of data parallelism in our implementation was 1.60 and that of model parallelism was 2.32. This indicates that model parallelism is faster than data parallelism for Seq2Seq RNN MT. We attribute this to the synchronization costs of a large number of parameters. The number of parameters used in the baseline model was 142 M and that for HybridNMT was 138 M.

Finally, the scaling factors of HybridNMTIF were between those of HybridNMT and the baseline model with model parallelism. This indicates that the proposed hybrid data-model parallel approach is faster than speed obtained when using only model parallelism, even when the same network structure is used. Furthermore, removing input-feeding allows for faster training speed.

\begin{figure}[t]
  \centering
  \includegraphics*[width=\columnwidth]{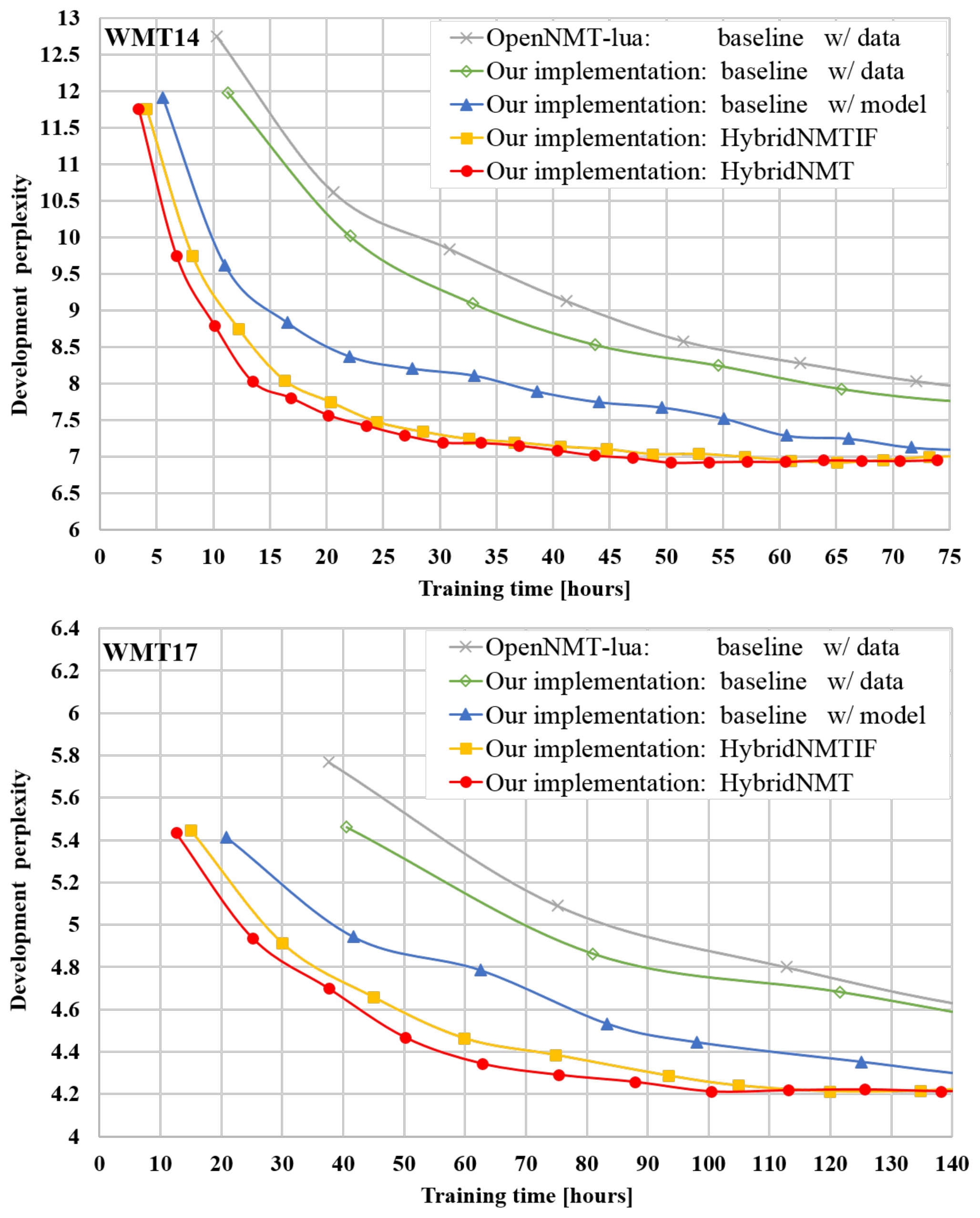}
  \caption{Convergence speed for different methods.}
  \label{fig:figure4}
\end{figure}

\subsection{Comparison of convergence speed}
\label{sec:comparison2}

\Fig{figure4} shows the convergence speed for different methods applied to WMT14 and WMT17. The horizontal axis represents wall-clock training time in hours. The vertical axis represents the perplexity of development data. We measured the perplexities at the ends of epochs, represented as points in the graphs.

HybridNMT converges faster compared with other methods. This, in addition to \Tab{table3}, implies that HybridNMT is better than other methods in terms of training and convergence speed. Other findings: data parallelism as implemented in both OpenNMT-lua and our implementation performed poorly as shown in \Fig{figure4} as well as in \Tab{table3}. The perplexities obtained with model parallelism became similar to those of our hybrid parallelism after long runs. Finally, the convergence speed of HybridNMTIF was between those of Hybrid-NMT and the baseline model with model parallelism. This indicates that the proposed hybrid data-model parallel approach is faster than model parallelism, and removing input-feeding leads to faster convergence.

\subsection{Translation accuracy}
\label{sec:accuracy}

As mentioned in \Sec{model}, the proposed Hybrid NMT uses a simpler model structure than that of the baseline model. We have shown in \Fig{figure4} that the perplexities of HybridNMT are comparable and even lower than those of the baseline model with data/model parallelism in a limited training time owing to its faster convergence speed. Herein, we compare the translation accuracy as measured by BLEU scores.

To compare BLEU scores, first, we selected the models for the proposed HybridNMT and OpenNMT-lua based on the information provided in \Fig{figure4}. In other words, we selected the models with the lowest development perplexities.

\Tab{table4} shows BLEU scores on the development data obtained by OpenNMT-lua and HybridNMT with diverse hyperparameters. The beam size was changed from 3 to 18. OpenNMT-lua used the same normalization method of GNMT \cite{Wu:16}. Its optimal parameters for the development data were as follows: the beam sizes were 6 and 12 for WMT14 and WMT17, respectively; the length normalization values were both 1.0; and the coverage normalization values were both 0. The proposed HybridNMT used the same normalization of Marian \cite{Junczys-Dowmunt:18}, which simply divided the model score using a length normalization factor. Its optimal parameters were as follows: the beam sizes were 15 and 12 for WMT14 and WMT17, respectively and the length penalties were 1.0 for both datasets, implying that the model score was divided by the number of target words to get the normalized score.

\begin{table*}[t]
  \small
  \centering
  \begin{tabular}{l|l|cc}
    \hline
\multirow{2}{*}{\textbf{System}} &\multirow{2}{*}{\textbf{Reference}} &\textbf{WMT14} &\textbf{WMT17}\\
&&\textbf{test2014}	&\textbf{test2017}\\
\hline
RNNsearch-LV	& \citet{Jean:15}	&19.4	&--\\
Deep-Att	& \citet{Zhou:16}	&20.6	&--\\
Luong	& \citet{Luong:15}	&20.9	&--\\
BPE-Char	& \citet{Chung:16}	&21.5	&--\\
seq2seq	& \citet{Britz:17}	&22.19	&--\\
OpenNMT-lua	& \citet{Klein:17}	&19.34	&--\\
	& Our experiment	&21.85	&25.92\\
HybridNMT	& Our experiment	&22.71	&26.91\\
GNMT	& \citet{Wu:16}	&24.61	&--\\
Nematus (deep model)	& \citet{Sennrich:17}	&--	&26.6\\
Marian (deep model)	& \citet{Junczys-Dowmunt:18}	&--	&27.7\\    \hline
  \end{tabular}
  \caption{BLEU scores published regarding Seq2Seq RNN MT.}
  \label{tab:table5}
\end{table*}

We measured BLEU scores for WMT14 and WMT17 test data using the parameters stated above. \Tab{table5} shows the BLEU scores together with other published results on the same test data using Seq2Seq RNN MT for reference. For the WMT14 dataset, the proposed HybridNMT outperformed all the others but GNMT \cite{Wu:16}. Note that GNMT used 8 layers for the encoder-decoder part, while the proposed HybridNMT used 4 layers. Note also that the BLEU score of OpenNMT-lua in this experiment was higher than that of \citet{Klein:17}. This is probably because \citet{Klein:17} used 2 layers but we used 4 layers in our experiments. For the WMT17 dataset, the proposed HybridNMT performed comparably with other results. The results show that the translation of the proposed HybridNMT is accurate comparably with other Seq2Seq RNN MT models.

\section{Conclusions}
\label{sec:conclusion}

We have proposed a hybrid data-model parallel approach for Seq2Seq RNN MT. We applied model parallelism to the encoder-decoder part and data parallelism to the attention-softmax part. The experimental results show that the proposed hybrid parallel approach achieved more than 4 times speed-up in training time using 4 GPUs. This is a very good result compared with data parallelism and model parallelism whose speed-up was around 1.6--1.7 and 2.3--2.5 times when the same 4 GPUs were used. We believe the proposed hybrid approach can also be applied to the Transformer translation model because it also has the encoder, decoder, and softmax layers.

\section*{Acknowledgments}

We would like to thank Atsushi Fujita and anonymous reviewers for their useful suggestions and comments in this paper.

\bibliographystyle{acl_natbib}
\bibliography{mtsummit2019}

\end{document}